\documentclass{aastex631}
\usepackage{amsmath}

\renewcommand{\textbf}[1]{#1}

\begin{document}

\title{Metallicity of the double Red Clump in the Milky Way Bulge}

\author[0009-0009-8060-697X]{Zofia Budzik}
\affiliation{Astronomical Observatory, University of Warsaw,\\
Al. Ujazdowskie 4, 00-478 Warsaw, Poland}

\author[0009-0004-6995-5611]{Matylda Łukaszewicz}
\affiliation{Astronomical Observatory, University of Warsaw,\\
Al. Ujazdowskie 4, 00-478 Warsaw, Poland}

\author[0000-0002-9245-6368]{Radosław Poleski}
\affiliation{Astronomical Observatory, University of Warsaw,\\
Al. Ujazdowskie 4, 00-478 Warsaw, Poland}

\correspondingauthor{Zofia Budzik}
\email{zbudzik@astrouw.edu.pl, rpoleski@astrouw.edu.pl}

\begin{abstract}
We present a reanalysis of the metallicity of the double red clump (RC) stars from the Milky Way bulge. Two leading explanations for the existence of the double RC concern possible differences in chemical composition (multiple populations of stars) or distance (X-shaped bulge). We aim to verify the chemical composition hypothesis by determining the mean metallicity of each RC. We use infrared photometric data and metallicities from a previous study (that are based on the Michigan/Magellan Fiber System spectra). In contrast with previous studies, we assign stars to RCs or the red giant branch using a Bayesian approach. Our resulting difference between mean metallicity of the two RCs equals: 
$0.23^{+0.19}_{-0.24}$ or $0.28^{+0.17}_{-0.18}~\mathrm{dex}$, if the red giant branch bump is included or not, respectively. Because of the high statistical error, the result does not confirm the multiple populations hypothesis.   
\end{abstract}


\section{Introduction} \label{sec:intro}
Galactic bulges are the central structures of spiral galaxies. The bulge of the Milky Way is the closest one and, hence, the one in which individual stars are resolved best. These stars are observed photometrically in multiple bands, which allows constructing color-magnitude diagrams (CMDs). Bulge CMDs show a tight grouping of red metal-rich stars from the horizontal branch called the red clump \citep[RC;][]{1994ApJ...429L..73S,2016ARA&A..54...95G}. The RC overlaps with the red giant branch (RGB), and in this part of the CMD, the individual stars cannot be uniquely assigned to RC or RGB based on single epoch photometry. The CMDs for selected sight-lines in the Milky Way bulge show a double RC, which was discovered in 2010 by \citet{McWilliam_2010} and \citet{Nataf_2010}.

Since then, two main hypotheses emerged to explain the existence of the double RC:
the X-shaped bulge \citep{McWilliam_2010, 2012ApJ...757L...7L, 10.1093/mnras/stt1376} or multiple populations of stars \citep{10.1093/mnras/stv1980, Joo_2017}. The former assumes the difference in distance, the latter assumes the difference in chemical composition. To verify these hypotheses, spectroscopic observations of the double RC stars were carried out by \citet{2021ApJ...907...47L}, \citet{2018ApJ...862L...8L}, \citet{2012A&A...546A..57U}, and \citet{2011ApJ...732L..36D}.
Analyses based on radial velocities of the double RC by \citet{2012A&A...546A..57U} and \citet{2011ApJ...732L..36D} confirm the X-shaped bulge hypothesis.
\citet{2018ApJ...862L...8L} found that stars from the two RCs show a stark difference in CN-band strength. \citet{2021ApJ...907...47L} concluded that the RCs vary in Fe abundance ($[\mathrm{Fe/H}]$-metallicity) by $0.149 \pm0.036 ~\mathrm{dex}$.  \citet{2021ApJ...907...47L} also took into consideration the presence of background RGB stars in the RC samples, reporting that only $\sim 27\%$ of assumed RC stars are genuine RC stars. Taking that into account \citet{2021ApJ...907...47L} claimed the metallicity difference of $ \sim 0.55~\mathrm{dex}$. Based on these results, they concluded that the RCs consist of stars from separate stellar populations.
\textbf{The X-shaped bulge hypothesis does not rule out a small difference in metallicity between the RCs, which can be caused by a metallicity gradient. The $[\mathrm{Fe/H}]$ decreases with increasing distance from the galactic plane by $\approx$ 0.45 dex per kpc \citep{2016PASA...33...22N}. For a field centred at  $(l \sim 1^\circ, b\sim -8^\circ)$, fainter RC is located at a distance of 8.8 kpc and brighter at 6.5 kpc \citep{McWilliam_2010}. Based on these distances the fainter RC is further from the galactic plane by 0.34 kpc and therefore should be more metal-poor. If the above $[\mathrm{Fe/H}]$ applies to the X-shaped structure, then we expect metallicity difference of 0.15 dex from a simple projection effect.}

Here, we aim to reanalyse existing metallicity measurements in order to calculate the mean metallicity of the two RCs and measure their difference.
Since one RC is brighter than the other and in the studied sample RGB stars are also present, to study the chemical composition of each group, stars need to be properly divided into these groups. \citet{2021ApJ...907...47L} used an arbitrary division by setting magnitude ranges for each group. In our analysis, we used a different technique. Our calculations follow Bayesian approach instead of a standard frequentist one, and rather than using a strict division, we calculate probabilities of belonging to each group.
We also take into account the RGB Bump (RGBB), a temporary increase in luminosity of RGB stars due to the passage of the hydrogen-burning shell through the composition discontinuity \citep{Nataf_2013,Miller_Bertolami_2023}.

The structure of this paper is as follows. Data are presented in Section 2. Sections 3 and 4 explain the analysis, and final results are included in Section 5.

\section{Data} \label{sec:style}
For this study, we used \textbf{three} datasets: a sample of spectroscopically measured metallicities from \citet{2021ApJ...907...47L} \textbf{and two} near-infrared photometric \textbf{catalogues:} Two Micron All Sky Survey \citep[2MASS;][]{2006AJ....131.1163S} \textbf{and VISTA Variables in the Vía Láctea Survey \citep[VVV;][]{2010NewA...15..433M}.}
Both datasets are centred on the bulge field at $(l \sim -1^\circ, b\sim -8.5^\circ)$ where observing the double RC is optimal.

\citet{2021ApJ...907...47L} chose around 450 targets based on their magnitudes in the $K$ filter from 2MASS. The survey was carried out with two 1.3 m telescopes located at the Mount Hopkins, Arizona, and Cerro Tololo, Chile. Both sites used cameras with HgCdTe, $256$ by $256$ pixels detectors with the pixel scale of $2''$.

For the chosen stars, \citet{2021ApJ...907...47L} observed spectra with resolution $ R\sim 21,000$ using the Michigan/Magellan Fiber System  \citep[M2FS;][]{2012SPIE.8446E..4YM} on the Magellan Clay 6.5 m Telescope at the Las Campanas Observatory, Chile. The spectra were reduced following the data reduction procedure of \citet{Johnson_2015}.
To obtain the metallicity, \citet{2021ApJ...907...47L} used standard spectroscopic methods \citep[see][]{refId0}. 
The uncertainties of $[\mathrm{Fe/H}]$ were calculated based on the abundances attained from each absorption line as well as systematic errors related to the atmospheres. After excluding spectra with too low signal-to-noise ratio, they ended up with a final sample of 354 stars.

\textbf{The VVV survey was conducted using a 4 m wide-field Visible and Infrared Survey Telescope for Astronomy (VISTA). The telescope is located at ESO’s Cerro Paranal Observatory in Chile. The near-infrared camera consists of pixels with mean size of  $0.34''$, which translates to a field of view diameter of $1.65^\circ$ \citep{2010NewA...15..433M}}

We used 2MASS and VVV catalogues to select two samples of stars from a field with the centre at  $(l \sim -1^\circ, b\sim -8.5^\circ)$ and radius of $22'$. We implemented a limit on declination $-34.2^\circ<\delta <-33.7^\circ$ to better match the \citet{2021ApJ...907...47L} sample. That yielded 26,129 stars from 2MASS \textbf{and 110,664 from VVV.} We further narrowed it down to stars from the RGB, namely stars with the $(J-K)$ colour in the range (0.5, 1.1) mag and $K$ magnitude in the range (11.55, 15) mag. We ended up with two samples of 10,406 and 9,888 stars.

\section{Bright and Faint Red Clump}
\subsection{Mixture model}
To calculate the mean metallicities of the RCs, selected targets need to be categorised into one of the three groups: RC1, RC2 (designated bRC and fRC in \citet{2021ApJ...907...47L}), and RGB. Instead of arbitrary division based on magnitudes, we calculated the probability that a given star belongs to either of the three groups. These probabilities are based on a mixture model that takes into account a number of stars from each group at a given brightness.
Mixture model is a probabilistic model that considers the existence of subgroups in the overall sample, without a direct measurement that indicates an affiliation to a certain group. For this case, the mixture model consists of two Gaussian distributions to represent the RCs:
\begin{equation}\label{rcgauss}
    f_{i}(K)= N_{i}\frac{1}{\sigma_{i}\sqrt{2\pi}}\exp\left(-\frac{(K-\mu_{i})^2}{2\sigma_{i}^2}\right)
\end{equation}
where $i=(RC1,RC2)$, $\mu_{i}$ is the mean value, $\sigma_{i}$ is standard deviation, and $N_i$ is the scaling factor.

To represent the RGB, we chose an exponential distribution (following the example of \citet{Nataf_2013}):
\begin{equation}
    f_{RGB}(K)=A\exp (BK).
\end{equation}
where $A$ is a scaling factor and $B$ is the slope parameter.

From these equations, we define the following probabilities that a star with brightness $K$ belongs to either of the three groups:

\begin{equation}\label{prob}
    P_{j}(K) = \frac{f_{j}(K)}{\sum_{j} f_j(K)}
\end{equation}
\begin{equation}
    \sum_{j} P_{j}(K) = 1   
\end{equation}
where $j=(RGB,RC1,RC2)$.

\subsection{RGBB}
We also examined a case where the RGBB is included in the mixture model. To account for the RGBB, we added another Gaussian distribution to the one described by Equation \ref{rcgauss}. The parameters of this distribution are described by parameters from Equation \ref{rcgauss} and the scaling factors derived by \citet{Nataf_2013} based on empirical fits of Galactic globular clusters:
\begin{equation}
\begin{aligned}
     N_{RGBB,i}=0.201 \times N_i \\
    \mu_{RGBB,i} = \mu_i + 0.737 \\
    \sigma_{RGBB,i} = \sigma_i.
\end{aligned}    
\end{equation}

\begin{figure}[ht!]
\begin{center}

\includegraphics[trim=1.5cm 0 1.5cm 0.5cm,width=1\textwidth]{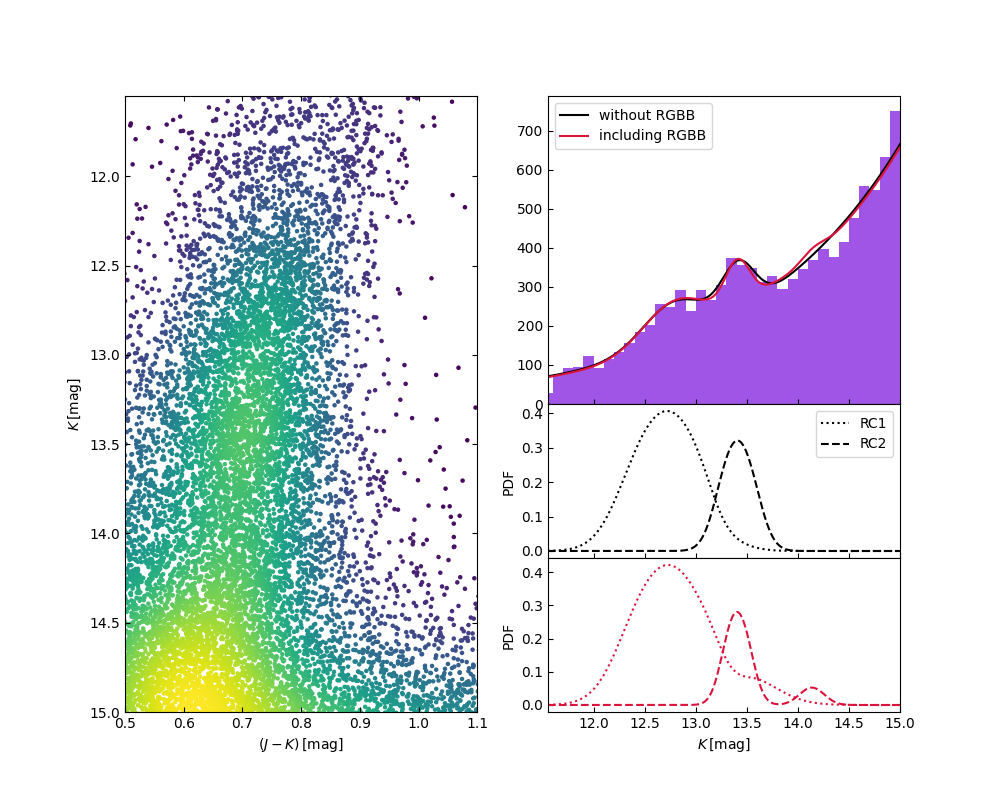}

\end{center}
\caption{CMD, the distribution of $K$-band brightness and PDFs of the two RCs.
All plots include 10,406 stars after the imposed colour and the $K$ magnitude restrictions. Warmer tones on the CMD indicate more densely populated regions.
The black and red lines are the best fits of the mixture model with or without RGBB. 
\label{fig:cmdhist}}
\end{figure}

\subsection{\textbf{Effective number of stars}}

\textbf{We estimated the effective number of stars in each group by summing probabilities: $ N_{eff,j =}\sum_{n} P_{j}(K_n)$, where $K_n$ is the brightness of a star with index $n$. The results of this calculations are presented in Table \ref{tabela2}.}

\begin{deluxetable}{cccccc}
\tablecaption{\textbf{Comparison of the effective number of stars in each group for different datasets}
\label{tabela2}
}

\tablehead{\colhead{} & \colhead{2MASS} & \colhead{2MASS} & \colhead{VVV} & \colhead{VVV} & \colhead{\citet{2021ApJ...907...47L}} \\ 
\colhead{} & \colhead{with RGBB} & \colhead{without RGBB} & \colhead{with RGBB} & \colhead{without RGBB} & \colhead{} } 

\startdata
RGB&248.6 & 251.0 & 245.9 & 249.0 & 24 \\
RC1&82.5 & 69.2 & 69.4 & 57.2 & 135 \\
RC2&22.9 & 33.8 & 38.7 & 47.8 & 164 \\
\enddata
\tablecomments{Columns correspond to different input catalogues and assumptions of RGBB. In \citet{2021ApJ...907...47L} the remaining 31 stars are in the twilight zone between one RC and the other.}
\end{deluxetable}

\section{Analysis}
\begin{figure}[ht!]
\begin{center}
    
\includegraphics[trim=0.5cm 0 0.5cm 0,width=\textwidth]{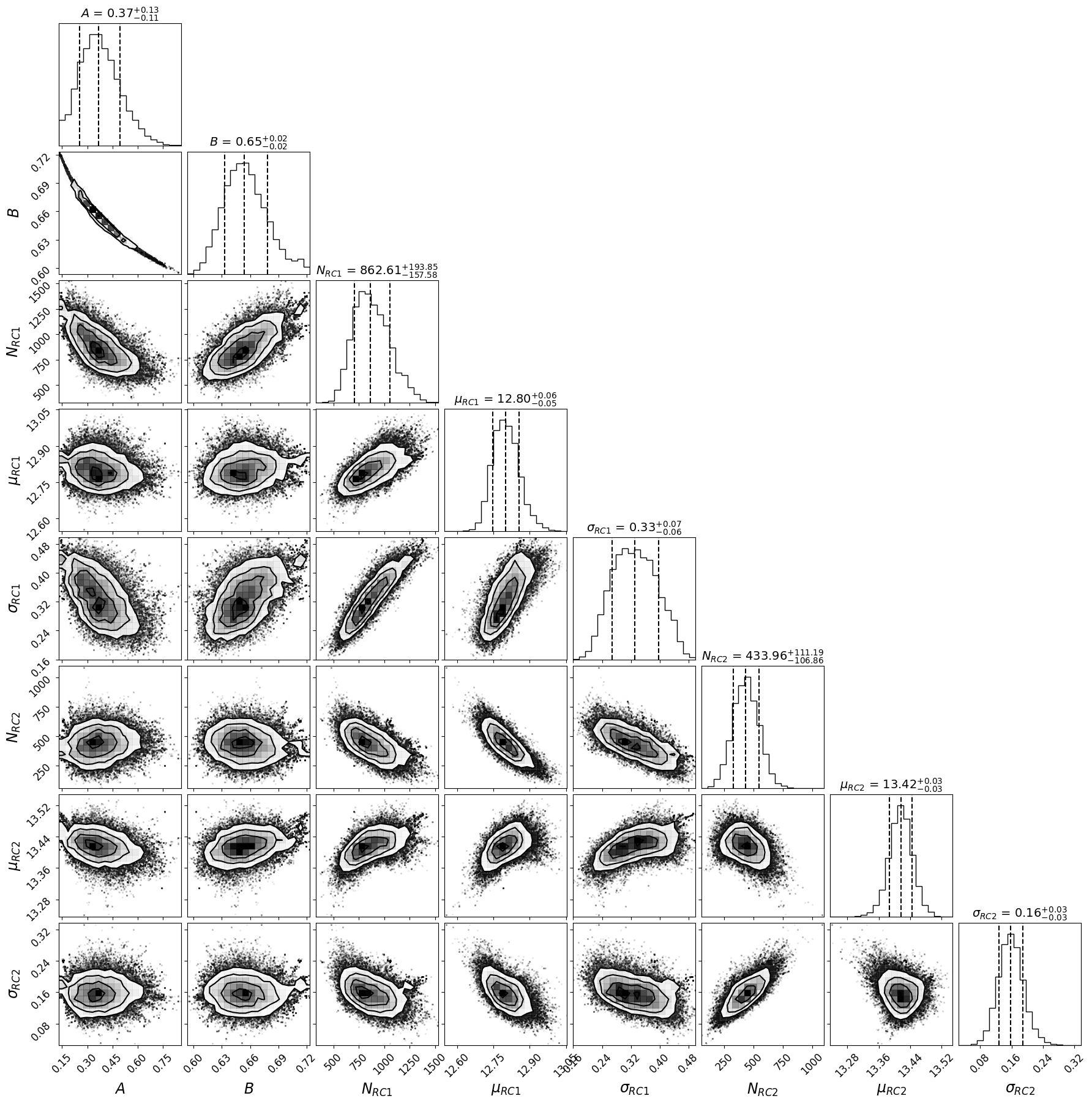}

\end{center}
\caption{Posterior distribution of fitted mixture model parameters without RGBB.\label{corner_None}}

\end{figure}

\begin{figure}[ht!]
\begin{center}
    
\includegraphics[trim=0.5cm 0 0.5cm 0,width=\textwidth]{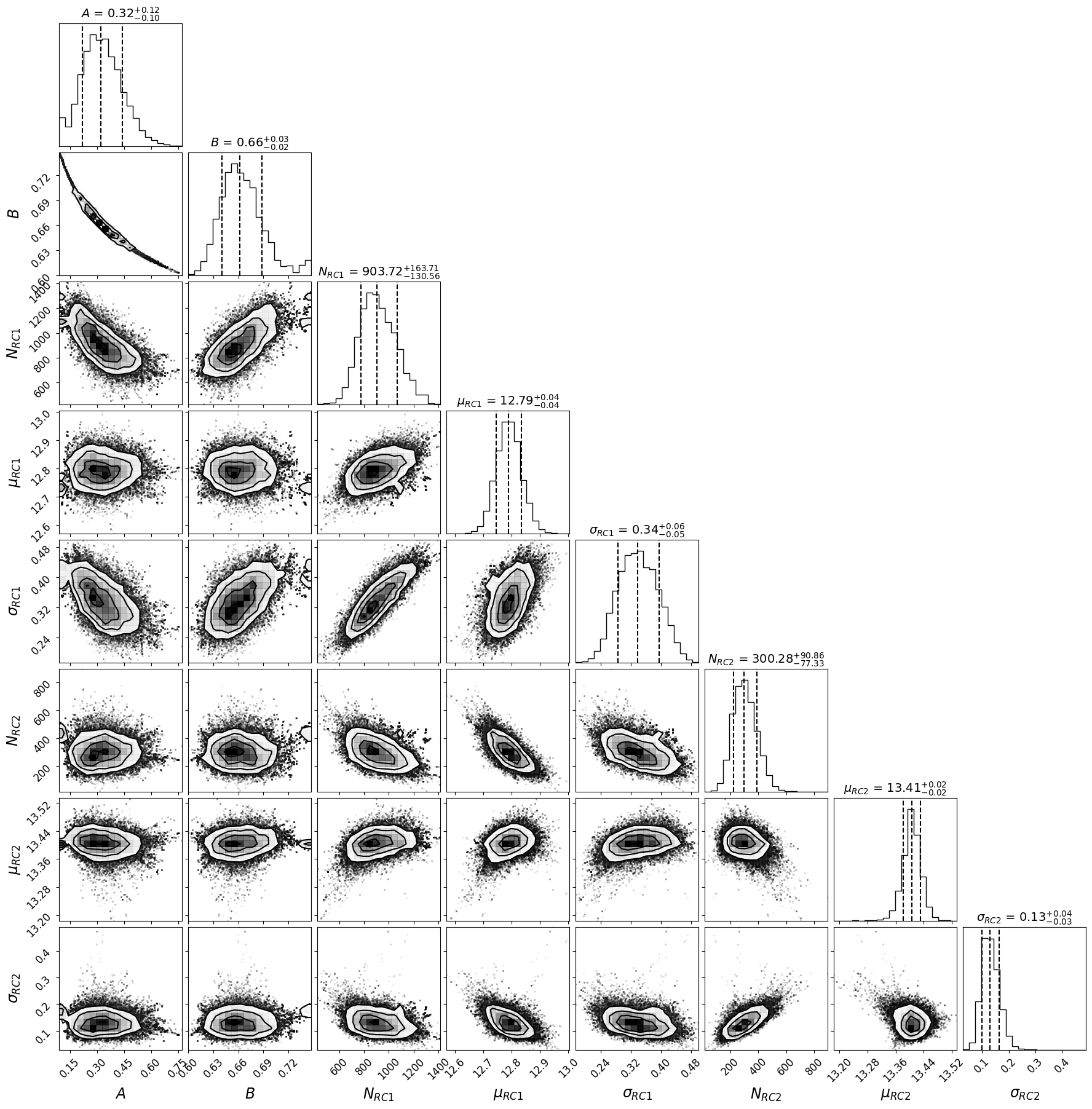}

\end{center}
\caption{Same as Figure \ref{corner_None} for the mixture model with RGBB.\label{corner_Nataf}}

\end{figure}

Parameters from the previous section together form a set of eight mixture model parameters: $(A, B, N_{RC1}, \mu_{RC1}, \sigma_{RC1}, N_{RC2}, \mu_{RC2}, \sigma_{RC2})$.
To acquire the most accurate parameters of the mixture model, we fitted the model to a sample of 10,406 stars.
To do that, we used \texttt{emcee} \citep{2013PASP..125..306F}, a Python implementation of the Monte Carlo Markov Chain ensemble sampler. The fitted model reflects the distribution in the upper right panel of Figure \ref{fig:cmdhist} well. We present the obtained posterior distributions of the parameters in Figures \ref{corner_None} and \ref{corner_Nataf} for the mixture models without and with RGBB, respectively.

To check the reliability of our fitting of mixture model parameters, we simulated samples of $K$ brightness assuming $maximum - a - posteriori$ results from fitting. Then we fitted the simulated data using the same procedure. The resulting posterior was similar to the ones in Figures \ref{corner_None} and \ref{corner_Nataf}. This check showed that our fitting method gives credible results.

In order to calculate the metallicity of each RC and subsequently the difference between them, we carried out another \texttt{emcee} fitting. We assumed that the metallicity distribution of each population is a normal distribution with mean $ \overline{[\mathrm{Fe/H}]}_j$, and dispersion $\sigma_j$. This time we determined: $ \overline{[\mathrm{Fe/H}]}_{RGB}, \sigma_{RGB}, \overline{[\mathrm{Fe/H}]}_{RC1}, \sigma_{RC1}, \overline{[\mathrm{Fe/H}]}_{RC2}, \sigma_{RC2}$ as the fitted parameters. With $maximum - a - posteriori$ parameters from the first fitting, we established probabilities $P_{RC1}(K), P_{RC2}(K), P_{RGB}(K)$ for every star in the final spectroscopic sample.
We used these as one of the arguments together with metallicities $[\mathrm{Fe/H}]_{n}$ and uncertainties $\epsilon_{n}$ derived by \citet{2021ApJ...907...47L}. The likelihood is given by the equation:
\begin{equation}
    \log\mathcal{L}=\sum_{n}\left(\log\left(\sum_{j} P_{j}(K_{n})\frac{\exp\left(-\frac{1}{2}\frac{\Delta_{nj}^2}{\sigma_{nj}^2}\right)}{\sqrt{2\pi\sigma_{nj}}}\right)\right),
\end{equation}
where 
\begin{equation}
    \sigma_{nj}^{2} = \epsilon_{n}^{2} + \sigma_{j}^{2}.
    \end{equation}
and 
\begin{equation}
    \Delta_{nj} = \overline{[\mathrm{Fe/H}]}_{j} - [\mathrm{Fe/H}]_{n}
\end{equation}

We set priors on mean $ \overline{[\mathrm{Fe/H}]}_j$, and dispersion $\sigma_j$. For mean values, we used a symmetric generalized normal distribution given by equation:
\begin{equation}
p(x)=\frac{\beta}{2\alpha\Gamma(1/\beta)}\exp\left({-\left(\frac{\left|x-\mu\right|}{\alpha}\right)^\beta}\right),
\end{equation}

where $\mu$ is location, $\alpha$ is scale, $\beta$ is shape, and $\Gamma$ is the gamma function.
We set them at $\mu=-0.5,\alpha=1,\beta=5$, which results in a top-flat distribution with a half maximum range from $-1.4$ to $0.4$.
 
For dispersion, we implemented a log normal distribution with scale and location set to $0.25$ and $0$, respectively, in order to reduce the existence of unlikely $\sigma>0.5$ points from the final distribution. The priors were chosen based on previous studies of the bulge metallicity \citep{2016PASA...33...22N,2018ARA&A..56..223B}.

\begin{figure}[ht!]
\begin{center}
\includegraphics[width=\textwidth]{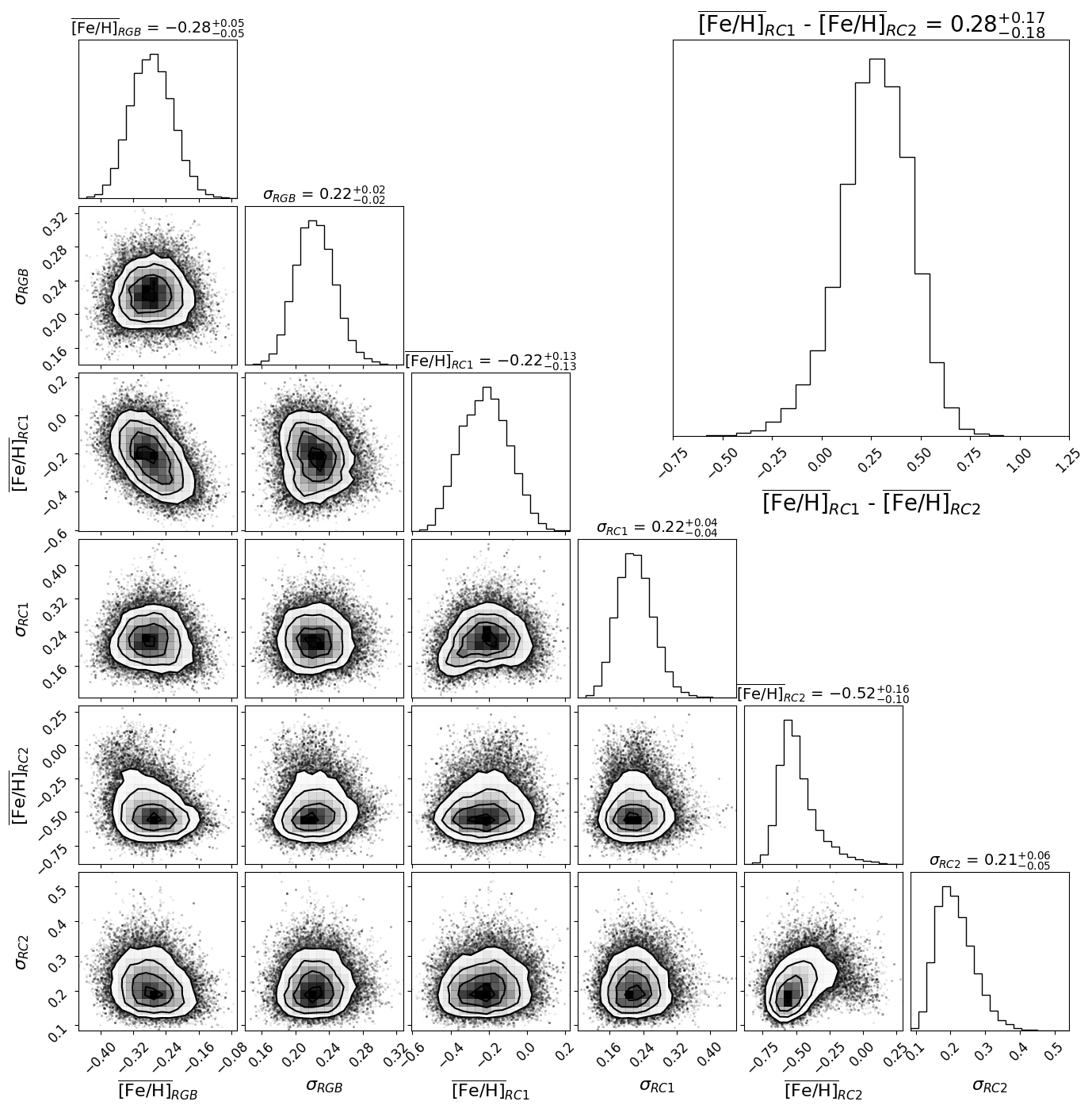}
\end{center}
\caption{Posterior distribution of fitted mean metallicities and their dispersions
for a mixture model without RGBB. The top right panel shows the posterior distribution of metallicity difference between two RC.\label{feh_None}}
\end{figure}

\begin{figure}[ht!]
\begin{center}
\includegraphics[width=\textwidth]{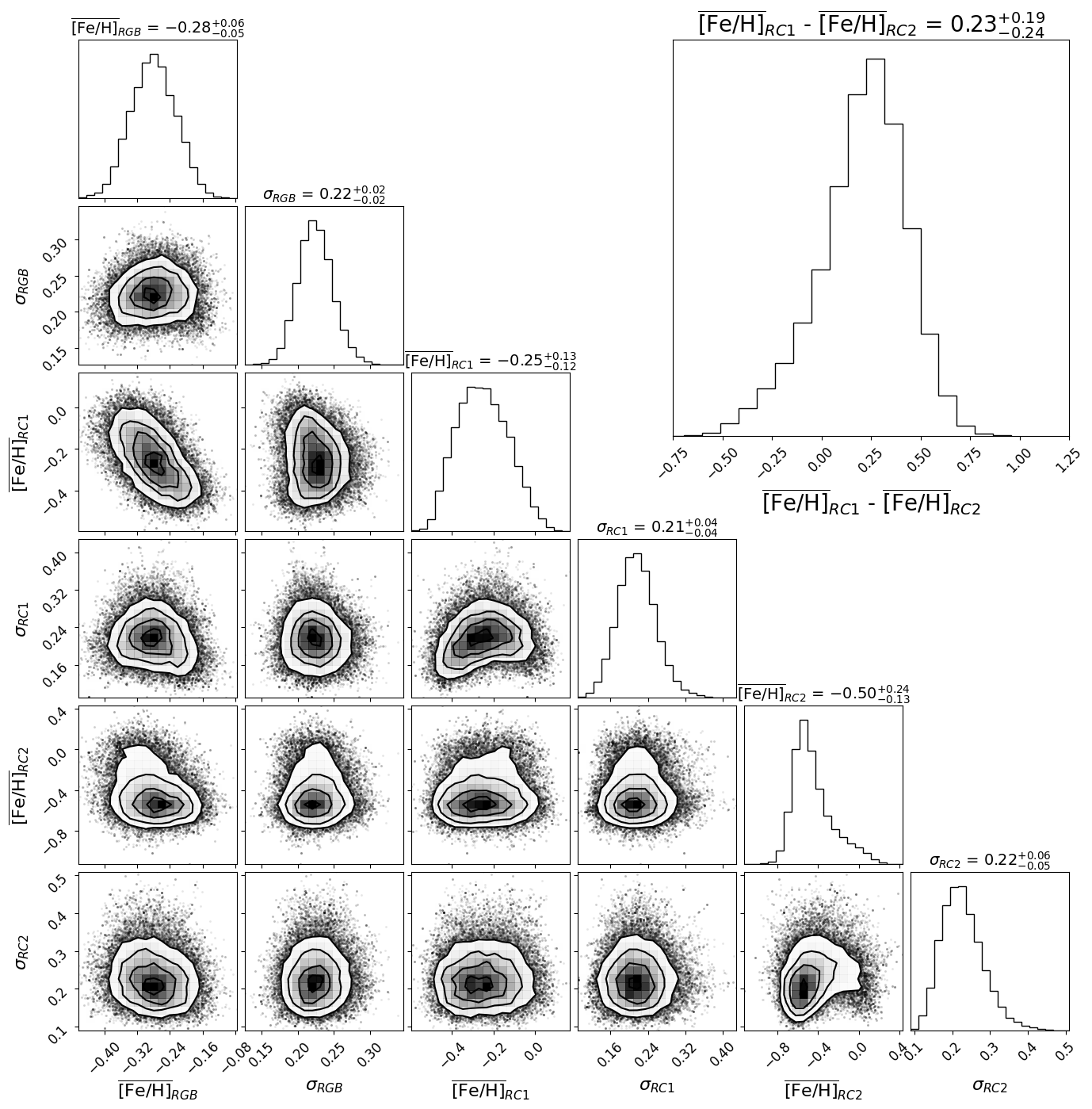}
\end{center}
\caption{Same as Figure \ref{feh_None} for the mixture model with RGBB.\label{feh_Nataf}}
\end{figure}

The final result of this fitting are the posterior distributions in Figure \ref{feh_None} and \ref{feh_Nataf}.

\clearpage

\subsection{\textbf{Results for the VVV data}}
\textbf{We also applied our framework to the VVV catalogue instead of 2MASS. The results are very similar, the resulting difference between mean metallicity of the two RCs equals: $0.27^{+0.16}_{-0.15}$ or $0.29^{+0.15}_{-0.14}~\mathrm{dex}$, if the red giant branch bump is included or not, respectively. One difference is the effective number of stars presented in Table \ref{tabela2}.}

\section{Results and Conclusions}
Based on our calculations, the mean difference in metallicity between the two RCs equals: $0.28^{+0.17}_{-0.18}~\mathrm{dex}$. Our result is subject to a high statistical error, so we can not definitely say that there is a significant difference.
Addition of the RGBB decreases the value of $N_{RC2}$ by a factor of 1.4 and decreases the mean metallicity difference to $0.23^{+0.19}_{-0.24}~\mathrm{dex}$. It also significantly changes the final distribution of the difference in metallicity between the two RCs. \textbf{These conclusions are also valid for the analysis based on the VVV data.}

Our reanalysis of \citet{2021ApJ...907...47L} metallicity measurements indicates no statistically significant difference in metallicity of the two RCs and is consistent with the X-shaped bulge hypothesis. The origin of the difference in metallicity in \citet{2021ApJ...907...47L} probably stems from the method of assignment to the groups.
Additionally, our analysis shows the importance of including the RGBB in further studies of the double RC.
The statistical framework presented in this paper can be used to better estimate the distribution of abundances using other samples, including abundances of single elements like sodium or aluminium.

\section*{Acknowledgments}
We are grateful to Andreas Koch-Hansen and\textbf{ an anonymous reviewer }for the constructive feedback that significantly contributed to improving the manuscript.
This research was funded in part by National Science Centre, Poland, grant SONATA BIS 2021/42/E/ST9/00038 awarded to R.P.
For the purpose of Open Access, the author has applied a CC-BY public
copyright license to any Author Accepted Manuscript (AAM) version arising from this submission.

\clearpage
\bibliography{sample631}{}
\bibliographystyle{aasjournal}

\end{document}